\documentclass[12pt]{iopart}
\usepackage{url}
\usepackage{hyperref}

\usepackage{xspace}
\usepackage{graphicx}% Include figure files

\newcommand{\bef}{\begin{figure}}
\newcommand{\eef}{\end{figure}}

\newcommand{\be}{\begin{equation}}
\newcommand{\ee}{\end{equation}}
\newcommand{\bea}{\begin{eqnarray}}
\newcommand{\eea}{\end{eqnarray}}
\newcommand{\sqsn}{\mbox{$\sqrt{s_{_{NN}}}$}\xspace}
\newcommand{\auau}{\mbox{Au$+$Au}\xspace}

\begin{document}

\title{Net-baryon number fluctuations with the hadron resonance gas model using Tsallis 
distribution}
\author{D.~K.~Mishra$^1$, P.~Garg$^2$, P.~K.~Netrakanti$^1$ and A.~K.~Mohanty$^1$}
\address{$^1$ Nuclear Physics Division, Bhabha Atomic Research Center, Mumbai 400085, India\\
$^2$ Department of Physics, Banaras Hindu University, Varanasi 221005, India}
\ead{dkmishra@rcf.rhic.bnl.gov}

\begin{abstract}
We explore a hadron resonance gas model using Tsallis non-extensive distribution to study the 
energy dependence of the product of the moments, $S\sigma$ and $\kappa\sigma^2$ of net-proton 
multiplicity distributions of published STAR data in \auau collisions at relativistic 
heavy-ion collider (RHIC) energies. 
While excellent agreements are found between model predictions and measurements of $S\sigma$ 
and $\kappa \sigma^2$ of most peripheral collisions and $S\sigma$ of most central collisions, 
the $\kappa \sigma^2$ for most central collisions deviates significantly from the predictions 
particularly at \sqsn = 19.6 GeV and $27$ GeV. This could be an indication of the 
presence of dynamical fluctuations, which are not contained in the HRG-Tsallis model. 

%\pacs{25.75.Bh \sep 25.75.Ld}
%      {PACS-key}{describing text of that key}   \and
%     {PACS-key}{describing text of that key}
%     } % end of PACS codes
\end{abstract}
\maketitle
\section{Introduction}
\label{intro}
The energy dependence of the moments (mean $M$, variance $\sigma$, skewness $S$ and kurtosis 
$\kappa$) and their products ($S\sigma$ and $\kappa\sigma^2$) for net-proton multiplicity 
distribution at RHIC energies are reported by STAR collaboration~\cite{Adamczyk:2013dal}. The 
product of the moments $S\sigma$ and $\kappa\sigma^{2}$ can be linked to the ratios of 
susceptibilities $(\chi)$ associated with the baryon number conservation 
\cite{Bazavov:2012vg,Aggarwal:2010wy,Gupta:2011wh}. For example, the product $S\sigma$ can be 
written as the ratio of third order ($\chi_B^3$) to second order ($\chi^2_B$) and the product 
$\kappa\sigma^{2}$ as the ratio of fourth order ($\chi_B^4$) to second order ($\chi_B^2$) 
baryon susceptibilities. The recent STAR measurements of $S\sigma$ and $\kappa\sigma^2$ show 
significant deviations from the predictions of the Skellam distribution (where $\kappa 
\sigma^2$ should be unity) at all energies, indicating the presence of large non-statistical 
fluctuations \cite{skellam}. The particle production in heavy ion collisions at relativistic 
energies are well described in terms of the hadron resonance gas (HRG) model where fermions and 
bosons follow Fermi-Dirac (FD) and Bose-Einstein (BE) distributions, 
respectively~\cite{Andronic:2005yp,Karsch:2010ck}. The success 
of HRG model would mean that the created thermal system which might have gone through a 
possible phase transition, has (nearly) equilibrated both thermally and chemically at 
freeze-out. It is 
believed that if the thermal system has retained some memory of the phase transition with 
finite correlation length at freeze-out, it must be reflected in the higher moments of the 
conserved quantities~\cite{Karsch:2010ck,Stephanov:2011pb,Cheng:2008zh}. Although it is not so 
obvious in the study of thermal abundance of the individual species. Therefore, the study of 
fluctuations in various conserved quantities such as: net-charge, net-strangeness and 
net-baryon number through the higher moments using HRG model is expected to provide a baseline 
to observe the deviation in experimental observables. The deviations of experimental data from 
these model studies may indicate the presence of non-statistical fluctuations, if any. 

The HRG model in Boltzmann approximation follows an exponential behavior of particle 
production corresponding to Boltzmann-Gibbs (BG) statistics. Recently, it has been shown that 
particle production both in heavy ion and proton-proton collisions at RHIC and large hadron 
collider (LHC) energies can 
be described successfully using a power law distribution at high transverse momentum rather 
than using the exponential one \cite{Tang:2008ud,Wong:2012zr,Cleymans:2011in}. Therefore, the 
Tsallis distribution function is being used for particle production with  non-extensive 
parameter $q$ such that it approaches Boltzmann distribution in the limit $q\rightarrow 1$. In 
the context of particle production in heavy ion collisions, Tsallis distribution has been 
interpreted as the superposition of Boltzmann distributions with different temperatures 
\cite{Wilk:1999dr}. Such a situation can occur when the hot and dense medium created in 
high-energy nuclear collisions is not homogeneous in temperature but fluctuates from point to 
point around some equilibrium value $T_f$ ~\cite{Wilk:2012zn,Stodolsky:1995ds}. The temperature 
fluctuation which exists in small part of the phase space with respect to the whole is another 
source of non-statistical fluctuation. This is different from the statistical fluctuations 
measured on an event by event basis and should be properly accounted in the model. In general, 
Tsallis non-extensive statistics is supposed to include situations characterized by long range 
interactions, long range microscopic memory and space time fractal structure of the 
process \cite{tsallis,Wilk:2006vp}. There could also be other non-statistical fluctuations
of dynamical nature associated with phase transition where the correlation length diverges
at the critical point \cite{Berdnikov:1999ph}. It may so happen that either due to finite size 
effect or if the freeze-out occurs at a temperature far away from the critical point, the 
strength of the dynamical fluctuations becomes progressively weaker. However, these dynamical 
fluctuations can still be studied through higher moments which diverge faster than the lower 
moment like the variance \cite{Stephanov:2008qz}.  

The success in understanding of non-equilibrated complex systems using Tsallis statistics 
\cite{tsallis,robledo} has motivated many phenomenological studies in understanding the 
particle production in elementary $p$+$p$ collisions at RHIC \cite{Adare:2010fe,Abelev:2006cs} 
and LHC \cite{Aamodt:2010my,Khachatryan:2010xs}. For heavy ion collisions, where the system 
formed is not a simple superposition of many $p$+$p$ collisions, Tsallis functional forms are 
successfully applied to describe the transverse momentum and rapidity distribution of the 
produced particles at different collision centralities and center of mass energies 
\cite{Tang:2008ud,Li:2014opa,Wilk:2008ue}. The non-extensive parameter $q$ characterizes the 
degree of non-equilibrium in the system. It indicates the deviations in the particle 
production mechanism from simple BG statistics and could give insight about the intrinsic 
fluctuations during hadronization process in the heavy-ion collisions \cite{Wilk:1999dr}.

In this paper, for the first time, we consider a hadron resonance gas model using 
Tsallis non-extensive distribution (HRG-Tsallis) to study non-statistical fluctuations of 
net-baryon (or net-proton) production in Au+Au collisions at RHIC energies. In the limit of 
$q\rightarrow 1$, we reproduce the HRG results. We show that the HRG-Tsallis model with a 
temperature dependent non-extensive $q$ parameter can explain the energy dependence of 
$S\sigma$ and $\kappa\sigma^2$ for most peripheral ($70 - 80\%$) collisions, but fail to 
explain the same for central ($0-5\%$) collisions. However, the energy dependence of 
$\kappa\sigma^2$ from experimental data for central collision deviates significantly 
from the HRG-Tsallis model predictions particularly at energies $19.6$ GeV and $27$ GeV. We 
argue that the predictions of HRG-Tsallis characterized by a temperature dependent $q$ 
parameter should be taken as the baseline to study (experimentally) fluctuations of dynamical 
origin if any, which are still not contained in the Tsallis non-extensive thermodynamics. 

The paper is organized as follows. In Section~\ref{sec:hrg}, we will discuss the HRG model 
with Tsallis distribution used in this study and interpretation of generalized susceptibility 
in terms of non-extensive parameter $q$. Section~\ref{sec:tvsq} describes the temperature 
dependence of $q$ parameter. In Section~\ref{sec:results}, comparison of the 
results for $\chi^{(3)}/\chi^{(2)}$ and $\chi^{(4)}/\chi^{(2)}$ between STAR experimental data 
for net-protons and HRG-Tsallis predictions for the net-baryon distributions for the most 
peripheral (70-80)$\%$ and the most central (0-5)$\%$ \auau collisions.
Finally in Section~\ref{sec:summary}, we summarize our findings and mention about the 
implications of this work to the current experimental measurements of higher moments in 
high energy heavy-ion collisions.
\section{HRG-Tsallis Model}
\label{sec:hrg}
The Tsallis form of FD and BE distribution can be written as \cite{Cleymans:2012ya},
\begin{equation}
f = \frac{1}{\mathrm{exp}_{q}\frac{(E-\mu)}{T}\pm 1}
\label{eqnt}
\end{equation}
where '$\pm$' signs are used for fermions and bosons, respectively and $\mathrm{exp}_{q}(x)$ 
is given by, 
\begin{equation}
\exp_q(x) = \left\{
\begin{array}{l l}
\left[1+(q-1)x\right]^{1/(q-1)}&~~\mathrm{if} ~~~x > 0 \\
\left[1+(1-q)x\right]^{1/(1-q)}&~~\mathrm{if}~~~x \leq 0 \\
\end{array} \right.
\label{eq:tsallis}
\end{equation}
where $x=(E-\mu)/T$. The above distribution approaches standard FD and BE distributions in the 
limit $q\rightarrow 1$.
Using the above non-extensive distribution function Eq.~\ref{eqnt}, 
the average number density can be written as,
\begin{eqnarray}
n_q &=&  \sum_i  g_{i} X_i\int\frac{d^3k}{(2\pi)^3} f_i^q(E_i,T_f,\mu_i),
\label{nq0} 
\end{eqnarray}
where $T_f$ is the chemical freeze-out temperature, $\mu_{i}$ is the chemical potential and 
$g_i$ is the degeneracy factor of the $i^{th}$ particle. The total chemical potential 
$\mu_{i}$ = $B_{i}\mu_{B}$ + $Q_{i}\mu_{Q}$ + $S_{i}\mu_{S}$, where $B_{i}$, $Q_{i}$ and 
$S_{i}$ are the baryon, electric charge and strangeness number of the $i^{th}$ particle, with 
corresponding chemical potentials $\mu_{B}$, $\mu_{Q}$ and $\mu_{S}$, respectively. The term 
$X_i$ represents either $B$, $Q$ or $S$ of the $i^{th}$ particle depending on whether the 
computed $n_q$ represents baryon density, electric charge density or strangeness density, 
respectively. The factor $d^3k$ can be expressed in terms of transverse momentum ($p_T$), 
pseudo-rapidity ($\eta$) and azimuthal angle ($\phi$) as, $d^{3}k=p_T\sqrt{p_T^2 + 
m^2}\cosh\eta~d{p_{T}} d\eta d\phi$ and energy $E$ is expressed as, $ E= \sqrt{p_T^2 + m^2} 
\cosh\eta$. Note that the exponent $q$ in $f_i$ has been introduced as a constraint for 
thermodynamical consistency~\cite{Cleymans:2012ya}. Since the average density 
and the pressure 
$P$ are shown to be thermodynamically consistent i.e. $n_q=\frac{\partial P_q}{\partial \mu}$, 
we can now define generalized susceptibilities as,
\begin{equation}
 \chi_{q}^{n} = \frac{\partial^n [P_{q}(T_f,\mu)]}{\partial \mu^n}|_T= \frac{\partial^{n-1} 
[n_q(T_f,\mu)]}
{\partial \mu^{n-1}}.
\label{eq:chi_q}
\end{equation}

Using Eq.~\ref{nq0} and Eq.~\ref{eq:chi_q}, we have calculated the susceptibility ratios 
$\chi^3/\chi^2$ ($\equiv S\sigma$) and $\chi^4/\chi^2$ ($\equiv \kappa\sigma^2$) for net-baryon 
distribution. We have also estimated moments of net-proton multiplicity distributions using 
only primordial protons, anti-protons as well as the yields coming from the resonance decays. 
The resonance corrections are carried out using an average efficiency as discussed in 
Ref.~\cite{Garg:2013ata}. We have noticed that within STAR acceptance, the differences between 
net-baryon and net-proton predictions are negligible. Therefore, in the present study we 
consider net-baryon distribution within STAR kinematic acceptance as used in 
Ref.~\cite{Adamczyk:2013dal}. We parametrized the freeze-out temperatures and chemical 
potentials using the relations, $\mu_B(\sqrt{s_{NN}}) $ = $\frac{d}{1+e\sqrt{s_{NN}}}$ and 
$T(\mu_B)$ = $a - b \mu_B^2 - c\mu_B^4$. For most central collisions, the parameters are taken 
from Ref.\cite{Karsch:2010ck,Cleymans:2005xv}.
% for most central collisions and are listed in Table~\ref{tab:mut_cent}. 
For peripheral collisions, we extracted these parameters from STAR 
preliminary data \cite{Das:2012yq} for most peripheral ($70-80\%$) centrality collisions and 
the extracted parameters are given in Table~\ref{tab:mut_peri}. We use similar parametrization 
for $\mu_S$ and $\mu_Q$ as that of $\mu_B$ and the corresponding parameters are also listed in 
Table~\ref{tab:mut_peri}. We set $\mu_Q$ to zero for peripheral collision as it does not have 
significant contribution as compared to $\mu_B$ and $\mu_S$.
%%%%%%%%TABLE-1%%%%%%%%%%%%%%
\begin{table}[h]
\caption{Parametrization of chemical potentials and freeze-out temperature extracted from 
the STAR experimental data for (70-80)\% centrality \cite{Das:2012yq}. }
\begin{center}
\begin{tabular}{lclclcl}
\hline
 & &     $a$~(GeV)           & &  $b$~(GeV$^{-1}$)     &      $c$~(GeV$^{-3}$)  & \\
\hline
T   &&  0.158 $\pm$ 0.002    & &  0.159 $\pm$ 0.034    & ~ 0.500 $\pm$ 0.001 & \\
\hline
\hline
 & & $d$~(GeV)   & &  $e$~(GeV$^{-1}$)     &       & \\
\hline
$\mu_{B}$   & &  0.900 $\pm$ 0.059            & &  0.251 $\pm$ 0.008&    &\\
$\mu_{S}$   & &  0.239 $\pm$ 0.001            & & 0.300 $\pm$ 0.001 &    & \\
\hline
\end{tabular}
\label{tab:mut_peri}
\end{center}
\end{table}

In the context of heavy ion collision, the Tsallis distribution has a simple interpretation in 
terms of the distribution of temperatures (instead of a single Boltzmann temperature) of the 
fire ball which is created during the collision process~\cite{Wilk:1999dr}. This can be shown 
mathematically as follows. The Tsallis-Boltzmann distribution as given in Eq.~\ref{eq:tsallis} 
can be written as,
\begin{equation}
\left (1+\frac{(E-\mu)}{T_f k}  \right )^{-k}=\int_0^\infty d\left (\frac{1}{T_B} 
\right )g \left( \frac{1}{T_B},\frac{1}{T_f}\right ) exp \left (-\frac{(E-\mu)}{T_B} \right ),
\end{equation}
where $k=1/(q-1)$ and $g$ is the gamma function given by,
\begin{equation}
g \left( \frac{1}{T_B},\frac{1}{T_f}\right )
=\frac{k T_f}{\Gamma(k)} {\left (\frac {k T_f}{T_B} \right )}^{k-1} exp \left 
( -\frac{k T_f}{T_B} \right ).
\label{eq:gamma}
\end{equation}

We assume a thermodynamic system constituting smaller systems which are 
described by a local temperature $T_B$, fluctuating from point to point around some 
equilibrium freeze-out temperature $T_f$.
The inverse temperature of smaller 
systems, ($1/T_{B}$), is gamma distributed around a mean,
\begin{equation}
\left <  \frac{1}{T_B} \right > = \frac{1}{T_f},
\end{equation}
and the parameter $q$ is related to the temperature fluctuation,
\begin{equation}
\frac{\left < \left ( \frac{1}{T_B}\right)^2\right>-\left <\frac{1}{T_B}\right > ^2 }{\left < 
\frac{1}{T_B} \right >^2}=q-1,
\end{equation} 
which becomes zero in the Boltzmann limit ($q \rightarrow 1$). This leads to the 
interpretation 
of the Tsallis distribution at $T_f$ as a superposition of Boltzmann distributions 
with different $T_B$. The parameter $q$ describes the spread around the average 
temperature $T_f$. 
Finally, the average baryon density can be written as, 
\begin{equation}
n_q(T_f,\mu)=\int_0^\infty d\left (\frac{1}{T_B} \right )
g \left( \frac{1}{T_B},\frac{1}{T_f}\right ) n^B(T_B).
\end{equation}
where $n^{B}(T_B)$ is the average baryon density obtained with Boltzmann statistics.
Using above relation, Eq.~\ref{eq:chi_q} can be expressed as weighted sum of the 
susceptibilities estimated using standard Boltzmann statistics and can be interpreted as an 
average taken over the whole phase space which is inhomogeneous in temperature. This is what 
will be measured experimentally if the temperature fluctuation exists in the hadronizing 
medium corresponding to a non-extensive value of $q$ which deviates from unity. Therefore, it 
is reasonable to argue that Eq.~\ref{eq:chi_q} can be used to estimate temperature averaged 
higher moments. It may be mentioned here that the above argument is strictly correct when 
quantum statistics is not important and Tsallis distribution can be written as the 
superposition of Boltzmann distributions. This condition is mostly true for baryons where quantum 
effects are small.

%%%%%%%%%%%%%%%%%%%%%%%Fig1%%%%%%%%%%%%%%%%%%%%%%%%%%%%%%
\begin{figure}[h]
\begin{center}
\includegraphics[width=1.0\linewidth]{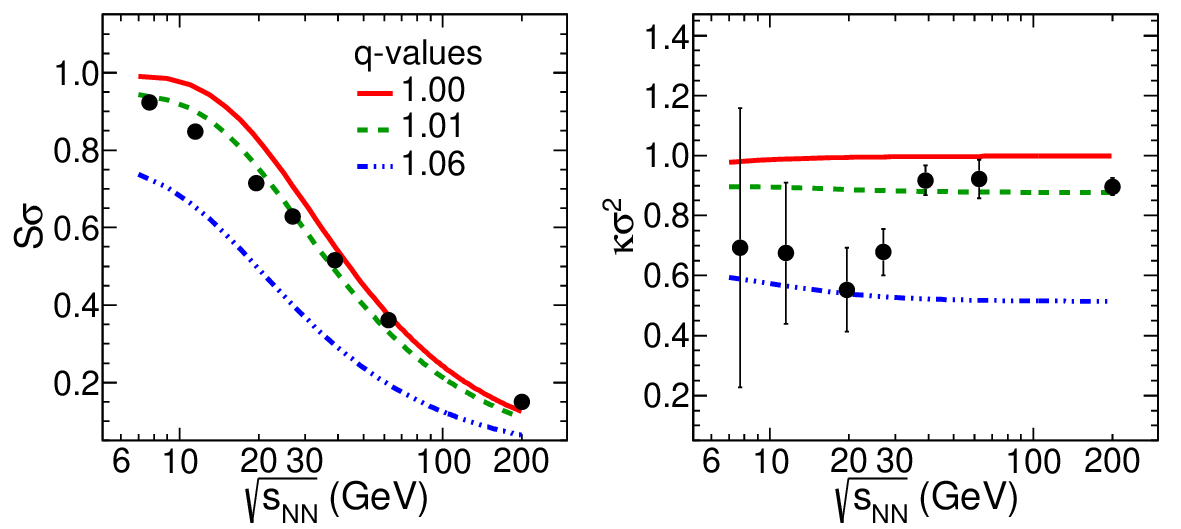}
 \caption{(Color online) The energy dependence of $S\sigma$ and $\kappa\sigma^2$ of net-baryon 
distribution calculated using HRG, where $q = 1$ (solid curve), HRG-Tsallis with $q=1.01$ 
(dashed curve) and $q=1.06$ (dashed-dotted curve). A fixed $q$ value is used for all 
collision energies. The filled circles are STAR data for most central ($0-5\%$) in Au+Au 
collisions \cite{Adamczyk:2013dal}.}
 \label{fig:fixq}
 \end{center}
 \end{figure}
%%%%%%%%%%%%%%%%%%%%%%%%%%%%%%%%%%%%%%%%%%%%%%%%%%%%%%%%%%

Figure~\ref{fig:fixq} shows the energy dependence of $S\sigma$ and $\kappa\sigma^2$ for 
central ($0-5\%$) \auau collisions for STAR data \cite{Adamczyk:2013dal}. Also shown are the 
predictions from HRG-Tsallis with different values of $q$. Higher value of $q$ indicates 
larger deviation from thermal equilibrium. For $q\rightarrow1$, the $S\sigma$ and 
$\kappa\sigma^2$ values approach the HRG prediction~\cite{Karsch:2010ck}. Both $S\sigma$ and 
$\kappa\sigma^2$ decrease with increasing values of $q$ for all collision energies. With $q$ = 
1.01, $S\sigma$ and $\kappa\sigma^2$ values from experimental data and HRG-Tsallis predictions 
are comparable at $\sqrt{s_{NN}}$ = 39, 62.4 and 200 GeV. However, for the agreements between 
data and HRG-Tsallis predictions for $\kappa\sigma^2$ at 19.6 and 27 GeV, we require a higher 
value of $q$ = 1.06. The larger value of $q$ at these two energies would indicate the need of 
energy dependent parametrization of $q$.

\section{Temperature dependent $q$ parameter}
\label{sec:tvsq}
The non-extensive $q$ parameter, which is generally extracted from the experimental data, 
assumes different values depending on the collision energy as well as on the centrality of 
the collisions. As discussed in Ref.~\cite{Tang:2008ud}, which uses transverse momentum 
spectra from the STAR experiment, the $q$ value decreases with increasing centrality 
indicating an evolution from a highly non-equilibrated system towards a more thermalized 
system. An increasing in centrality would mean higher temperature and lower baryo-chemical 
potential. At higher temperature, the system is expected to be closer to the equilibrium 
and gradually starts deviating from equilibrium as collision energy decreases or the collision 
becomes more peripheral. More specifically, $q$ should depend on both temperature $T_f$ and 
chemical potential $\mu_B$ at freeze-out. In this work, however, we adopt a simple approach and 
allow $q$ to depend only on $T_f$. On the other hand, $T_f$ is estimated from $\mu_B$ through a 
parametrization depending on the collision energy and the centrality of the collision as 
discussed before. Therefore, following Ref.~\cite{Tang:2008ud}, we express $q$ using the 
relation, $q = 1 + [\alpha(T_0 - T_f)]^{1/2}$, where $T_{0}$ is a reference temperature close 
to the freeze-out temperature at $\sqrt {s_{NN}}=200$ GeV in central ($0-5$\%) collisions and  
is fixed at $0.167$ GeV such that $q=1$ for $T_f \geq T_0$. 
%%%%%%%%%%%%%%%%%%%Fig2%%%%%%%%%%%%%%%%%%%%%%%%%%%%%%%%%%%%%
 \begin{figure}[htb]
\begin{center}
\includegraphics[scale=0.55]{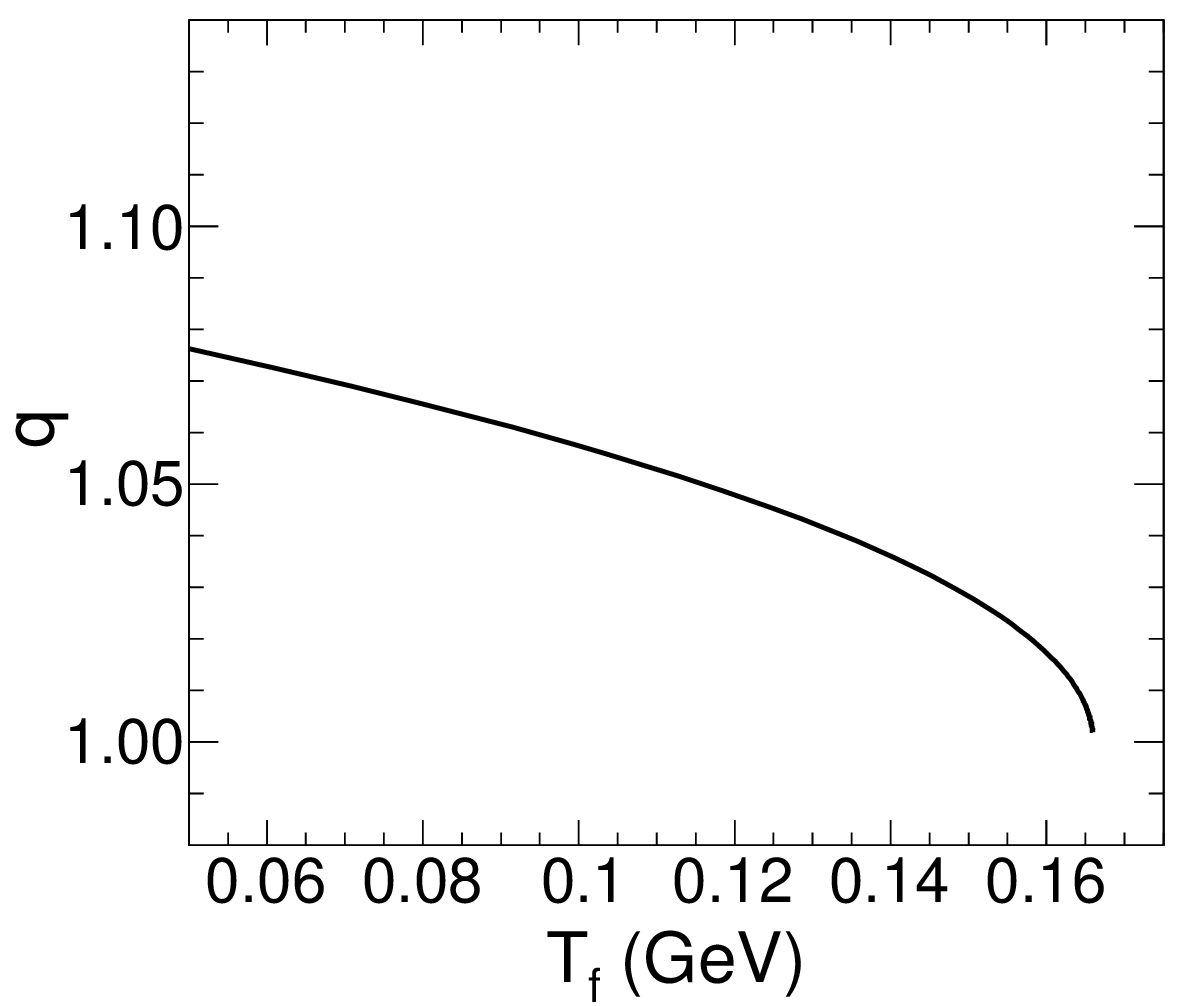}
 \caption{ The non-extensive parameter $q$ as a function of temperature corresponding to 
different $\sqrt{s_{NN}}$ for $\alpha = 0.05$.}
 \label{fig:tvsq}
 \end{center}
 \end{figure}
%%%%%%%%%%%%%%%%%%%%%%%%%%%%%%%%%%%%%%%%%%%%%%%%%%%%%%%%%%%%%

Figure~\ref{fig:tvsq} shows the temperature dependence of non-extensive parameter $q$. The 
$q$-values increase with decreasing $T_f$, which would mean that, the deviation of the 
hadronizing system from the equilibrium increases as collision energy decreases or as the 
collision becomes more peripheral. The $q$ value approaches unity at \sqsn = 200 GeV in 
central collisions. 
The parameter $\alpha$ is kept free and adjusted to fit the data. When $\alpha=0$ 
($q\rightarrow 1$), we get back the HRG results. Note that we use the same parametrization 
irrespective of whether the collision is central or peripheral except different centrality 
which will have different chemical freeze-out parameters. 

The partonic system, which is formed at high collision energy, has sufficient time between 
formation and freeze-out to achieve full equilibration, which corresponds to the region of 
$q=1$. This may not be the case at lower collision energies. The time available between 
formation and freeze out may not be sufficient to drive the system into the full 
equilibration. It could so happen that the hadronizing medium 
has attained only local equilibrium with different temperatures prevailing at different 
regions which can be considered as the fluctuations around some mean temperature $T_f$. This 
corresponds to the region of $q$ deviating from unity. However, by decreasing collision energy 
further, it may not be possible to produce the partonic system any more and the collisions may 
remain fully hadronic. 

%%%%%%%%%%%%%%%%%%%%%Fig3%%%%%%%%%%%%%%%%%%%%%%%%%%%%%%%%%
\begin{figure}[htb]
\begin{center}
\includegraphics[width=1.0\linewidth]{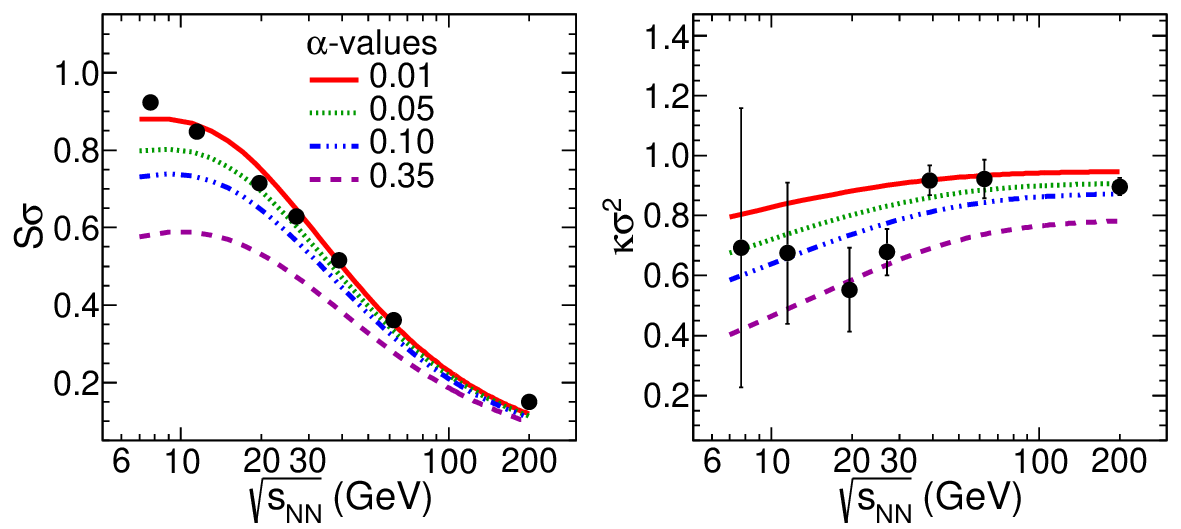}
 \caption{(Color online) The energy dependence of $S\sigma$ and $\kappa\sigma^2$ of net-baryon 
distribution calculated using HRG-Tsallis with different $\alpha$ values. The non-extensive 
parameter $q$ is related to $\alpha$ as described in the text varied with collision energy. 
The 
filled circles are STAR data for most central ($0$-$5\%$ centrality) in \auau collisions 
\cite{Adamczyk:2013dal}.}
 \label{fig:varq}
 \end{center}
 \end{figure}
%%%%%%%%%%%%%%%%%%%%%%%%%%%%%%%%%%%%%%%%%%%%%%%%%%%%%%%%%%
\section{Results and Discussions}
\label{sec:results}

Figure~\ref{fig:varq} shows the comparisons of $S\sigma$ and $\kappa\sigma^2$ between 
experimental data and HRG-Tsallis predictions with $q$ parametrization as shown 
in Fig.~\ref{fig:tvsq}. For $S\sigma$ and $\kappa\sigma^{2}$ at $\sqrt{s_{NN}}$ = 39, 62.4 and 
200 GeV, HRG-Tsallis predictions with lower value of $\alpha$ = 0.01 (related to $q$), 
describes the experimental data very well. However, the $\kappa\sigma^{2}$ values at 
$\sqrt{s_{NN}}$ = 19.6 and 27 GeV would require larger value of $\alpha$ ($\sim$ 0.35) to 
match with the experimental data. This further emphasizes the strong energy dependence of 
non-extensive parameter $q$ at these particular energies, indicating larger deviations from 
the thermal equilibrium. We study the sensitivity of HRG-Tsallis results for $S\sigma$ and 
$\kappa\sigma^2$ by comparing the experimental data with model for peripheral ($70-80\%$) and 
central ($0-5\%$) collisions, using the $q$-parametrization.

%%%%%%%%%%%%%%%%%%%%%%%%%%%Fig4%%%%%%%%%%%%%%%%%%%%%%%%%%%%%%%
\begin{figure}[hb]
 \begin{center}
\includegraphics[width=1.0\linewidth]{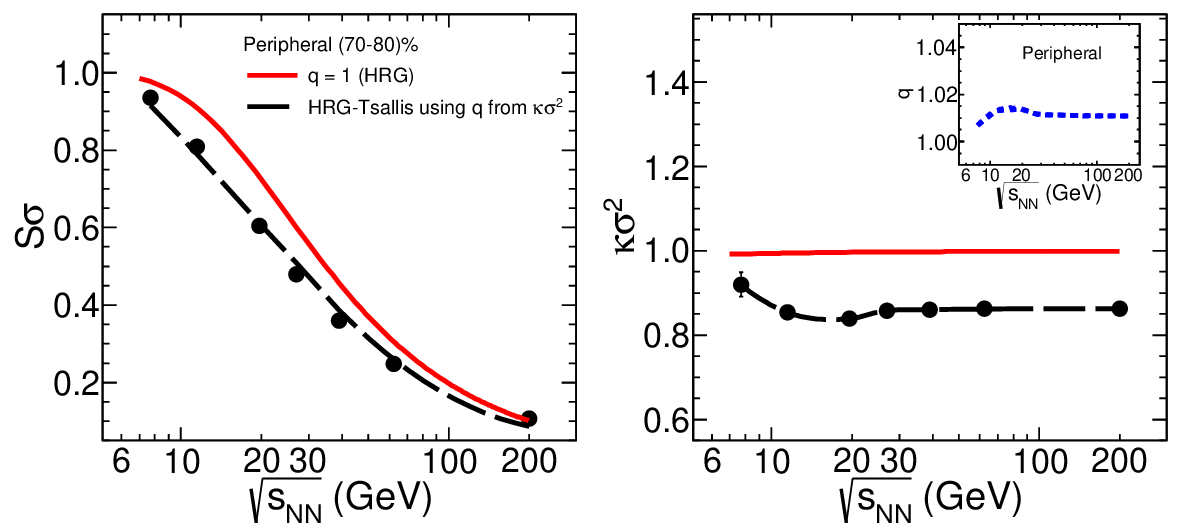}
\caption{(Color online) The energy dependence of $S\sigma$ and $\kappa\sigma^2$ of net-baryon 
distribution calculated using HRG (solid curve) and HRG-Tsallis with different $q$ 
values. The filled circles are STAR data for most peripheral ($70-80\%$ centrality) in 
Au+Au collisions~\cite{Adamczyk:2013dal}. The inset in right panel shows the variation of $q$ 
as a function of 
collision energy.}
 \label{fig4}
 \end{center}
 \end{figure}
%%%%%%%%%%%%%%%%%%%%%%%%%%%%%%%%%%%%%%%%%%%%%%%%%%%%%%%%%%%%%%%
%%%%%%%%%%%%%%%%%%%%%%%%%Fig5%%%%%%%%%%%%%%%%%%%%%%%%%%%%%%%%
\begin{figure}[htb]
\begin{center}
\includegraphics[width=1.0\linewidth]{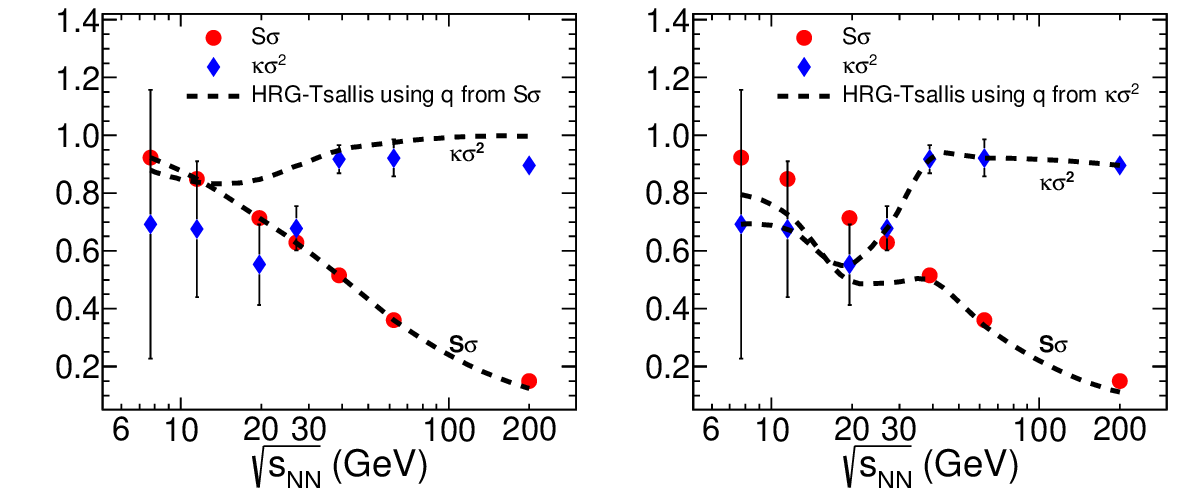}
 \caption{(Color online) The energy dependence of $S\sigma$ and $\kappa\sigma^2$ of net-baryon 
distribution calculated using HRG-Tsallis with different $q$ values. Left panel shows the 
HRG-Tsallis calculations which reproduce the experimental $S\sigma$ data and right panel 
shows the HRG-Tsallis calculations which reproduce the experimental $\kappa\sigma^2$ data. The 
symbols are STAR data for most central ($0-5\%$ centrality) in Au+Au collisions 
\cite{Adamczyk:2013dal}.}
 \label{fig5}
 \end{center}
 \end{figure}
%%%%%%%%%%%%%%%%%%%%%%%%%%%%%%%%%%%%%%%%%%%%%%%%%%%%%%%%%%%%%%%%%%

Figure~\ref{fig4}, shows the energy dependence of $S\sigma$ and $\kappa\sigma^2$ estimated 
using the freeze-out parameters as listed in Table~\ref{tab:mut_peri} for peripheral 
($70-80\%$) collisions. As can be seen, the HRG results ($\alpha$ = 0) show 
significant deviations from the experimental values~\cite{Adamczyk:2013dal}, and 
$\kappa\sigma^2$ values in HRG model is always close to unity where the data points are about 
$15\%$ below the HRG values. On the other hand, HRG-Tsallis with an average $q$ $\sim$ 1.01 can 
explain both $S\sigma$ and $\kappa\sigma^2$ very well for all the collision energies. The inset 
in Fig.~\ref{fig4} right panel shows the $q$ values used in HRG-Tsallis model to obtain a good 
agreement between experimental data of $\kappa\sigma^2$ and model results.

%%%%%%%%%%%%%%%%%%%%%%%%%Fig6%%%%%%%%%%%%%%%%%%%%%%%%%%%%%%%%
\begin{figure}[hb]
\begin{center}
\includegraphics[scale=0.55]{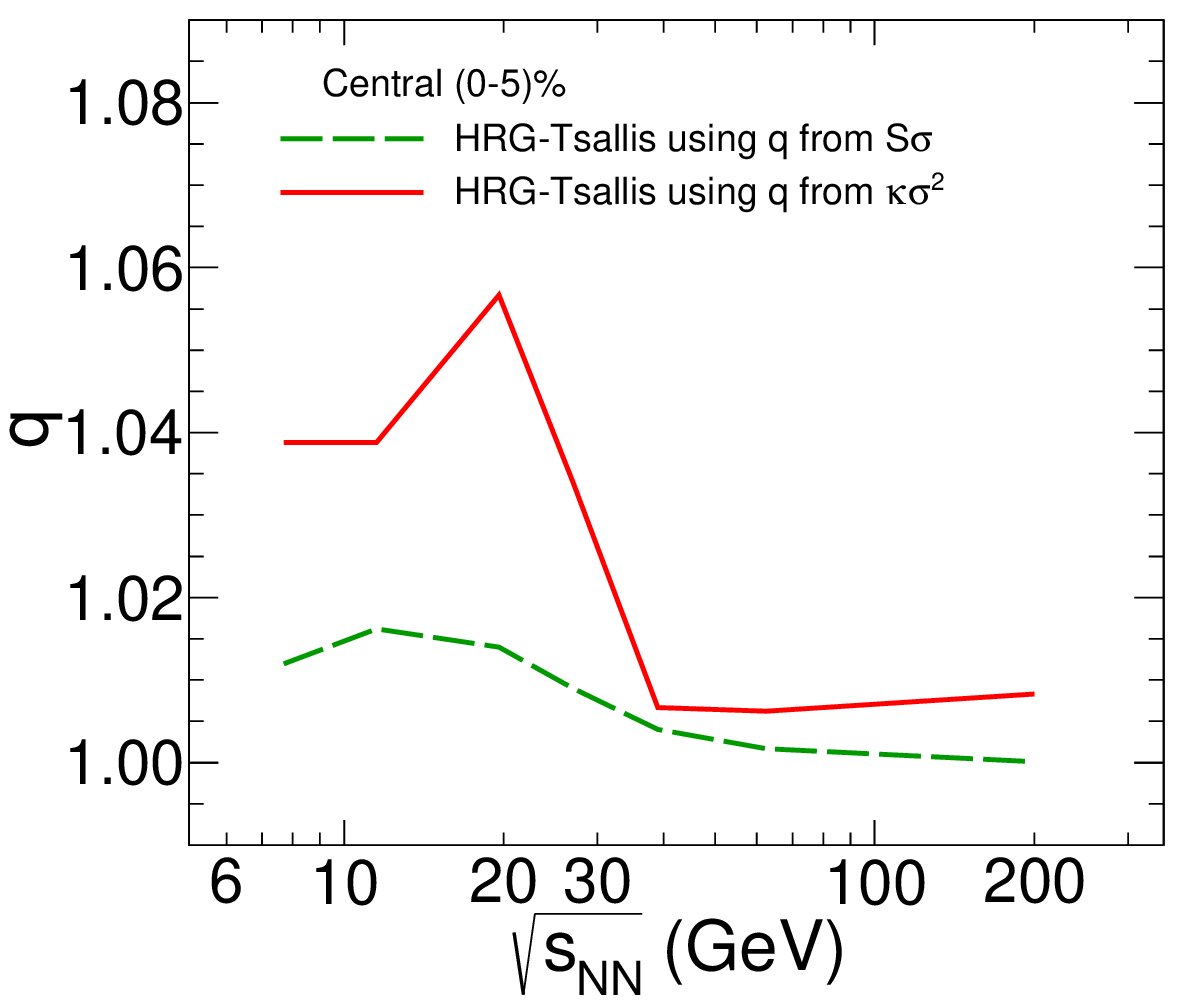}
 \caption{(Color online) The energy dependence of non-extensive parameter $q$ using 
HRG-Tsallis for central ($0-5\%$) collisions. The dashed curve shows the $q$ values obtained 
from reasonable agreement of $S\sigma$ between experimental and HRG-Tsallis results. The solid 
curve shows the $q$ values obtained from reasonable agreement of  $\kappa\sigma^2$ between 
experimental and HRG-Tsallis results.}
 \label{fig6}
 \end{center}
 \end{figure}
%%%%%%%%%%%%%%%%%%%%%%%%%%%%%%%%%%%%%%%%%%%%%%%%%%%%%%%%%%%%%%%%%%
The corresponding results for most central ($0-5\%$) centrality collisions are shown in 
Fig.~\ref{fig5}. The HRG predictions do not agree well the experimental data. 
We study the deviation of experimental data from the HRG results using two different methods.
In first method, we reproduce the $S\sigma$ values by taking different $q$ values ranging 
between 1.0 to 1.015 in the HRG-Tsallis model. Figure~\ref{fig5} left panel shows the 
comparison of experimental data and the HRG-Tsallis results (dashed curve) for $S\sigma$ and 
$\kappa\sigma^2$ obtained from the first method. The experimental data for $S\sigma$ at all 
energies nicely agrees with the HRG-Tsallis with energy dependent $q$ values, but fails to 
explain the $\kappa\sigma^2$ values particularly at lower energies (\sqsn $<$ 39 GeV). In 
second method, we reproduce the experimentally measured $\kappa\sigma^2$ data using different 
$q$ values ranging from 1.0 to 1.06 in the HRG-Tsallis model. Figure~\ref{fig5} right panel 
shows the comparison of experimental data and the HRG-Tsallis results (dashed curve) for 
$S\sigma$ and $\kappa\sigma^2$ obtained from the second method. Although using higher values 
of $q$ can explain the energy dependence of $\kappa\sigma^2$ at lower energies, but fails to 
explain the $S\sigma$ experimental data as shown in the right panel of Fig.~\ref{fig5}. The 
agreements between experimental data and HRG-Tsallis model for $S\sigma$ and $\kappa\sigma^2$, 
using a variable $q$ values, are difficult to obtain simultaneously under the current 
framework. The energy dependence of $q$ for ($0-5\%$) central collisions are shown in 
Fig.~\ref{fig6} for 
the above discussed methods. For central collisions, using $q$ values obtained from reasonable 
agreement between HRG-Tsallis and experimental data for $S\sigma$, there is smooth increase of 
$q$ ranging from 1.0 to 1.015 as we go from higher energies to lower energies. In second 
method, where we obtain $q$ values from reasonable agreement between HRG-Tsallis and the 
experimental measured $\kappa\sigma^2$ values, there is a sudden increase of $q$ parameter 
from 1.0 to 1.06 for energies below 39 GeV. Higher values of $q$ indicates larger deviation of 
the system from thermal equilibrium. Similar observation is also reported in 
Ref.~\cite{Cleymans:2008mt}.

It may be mentioned here that, we have not considered the Van der Waals (VDW) type excluded 
volume effect which could be an another source of deviation of $S\sigma$ and $\kappa\sigma^2$ 
from the HRG predictions~\cite{Fu:2013gga}. However, the VDW type deviation increases with 
decreasing energies where as the STAR experimental values have maximum deviation at 
$\sqrt{s_{NN}}=19.6$ GeV and less deviation at the two lower RHIC energies with larger 
uncertainties. Therefore, we have not considered VDW type effect in the present calculation. 
The transport model like UrQMD within STAR acceptance also produces similar results for 
$S\sigma$ and $\kappa\sigma^2$ which decrease with decreasing energies.
Tsallis distribution also includes quantum effect which is important at lower collision 
energies. Therefore, in this work we explore, how much non-statistical fluctuations inherently 
present in the Tsallis non-extensive distribution which can explain the experimental 
observations without considering other dynamical effects. 

Another aspect which has not been considered in current HRG-Tsallis model is the effect of 
non-extensivity on the freeze-out parameters which are generally extracted from the experiments 
using HRG model in grand canonical ensemble. It is observed in Ref.~\cite{Cleymans:2008mt} 
that while freeze-out temperature decreases, chemical potential increases with increasing $q$ 
parameter. However, for $q<1.01$ which has been used in the present study, we notice that the 
increase in $\mu_B$ is not significant although $T_f$ decreases by about $10\%$ from the value 
when $q=1$. As argued in~\cite{Cleymans:2008mt}, since Tsallis distribution is broader than 
the Boltzmann distribution, temperature needs to be decreased in order to  conserve the 
particle density. Therefore, we have estimated the moments keeping $\mu_B$ unchanged but 
allowing  freeze-out temperature to decrease up to  $10\%$. Interestingly, we notice that while 
$S\sigma$ increases slightly, $k\sigma^2$ remains unchanged. This suggests that $S\sigma$ is 
sensitive to both temperature and chemical potential while $\kappa \sigma^2$ is more sensitive 
to $q$ parameter.
\section{Summary}
\label{sec:summary}
In conclusion, we have studied the energy dependence of the fluctuations of net-baryon 
productions through higher moments namely $S\sigma$ and $\kappa\sigma^2$ using HRG with Tsallis 
non-extensive distribution function. When the non-extensive parameter $q$ is close to unity, 
the moments obtained using HRG-Tsallis model can be interpreted as the weighted average of the 
moments estimated using many Boltzmann distributions corresponding to the distribution of 
temperatures over the whole phase space. It is shown that HRG-Tsallis model can explain the 
energy dependence of $S\sigma$ and $\kappa\sigma^2$ measurements of the most peripheral 
collisions which is otherwise difficult to explain using the normal HRG model which predicts 
$\kappa\sigma^2$ close to unity for net-baryon productions. The HRG-Tsallis model also explains 
the energy dependence of $S\sigma$ data of central collision. However, the model can not 
explain the corresponding $k\sigma^2$ values for the central collisions particularly at 
energies $19.6$ GeV and $27$ GeV at the same time. This deviation is so significant that it is 
an indication of the presence of additional fluctuations at around $20$ GeV which may have 
some dynamical origin not contained in the HRG-Tsallis model. These dynamical fluctuations may 
originate from a QCD phase transition of hadronic to partonic medium. This study also provides 
an alternative baseline for the experimental results for the moments of the net-proton 
multiplicity distributions which may indicate the possible presence of a phase transition and 
the critical point in high energy heavy-ion collisions.

\noindent {\bf Acknowledgments }\\
Financial assistance from the Department of Atomic Energy, Government of India is
gratefully acknowledged. PG acknowledges the financial support from CSIR, New Delhi, India.
\\
%
% BibTeX users please use
% \bibliographystyle{}
% \bibliography{}

\begin{thebibliography}{}
\bibitem{skellam}
The Skellam distribution is the discrete probability distribution
of the difference $N_1-N_2$ where $N_1$ and $N_2$ are
two random variables each follwing Poisson distribution.
\end{thebibliography}
%
% Non-BibTeX users please use

\end{document}